\documentclass[draft,showpacs,preprintnumbers,amsmath,amssymb]{revtex4}
\usepackage[utf8]{inputenc} 
\usepackage{array}
\begin{document}

\title{Higher Dimensional Cylindrical or Kasner Type   
Electrovacuum Solutions}

\author{\"Ozg\"ur Delice}
\email{ozgur.delice@marmara.edu.tr}
\affiliation{Physics Department, Marmara University, Faculty of Science and Letters, Istanbul, Turkey}

\author{P\i nar Kirezli}
\email{pkirezli@nku.edu.tr}

\author{Dilek K. \c{C}iftci}
\email{dkazici@nku.edu.tr}
\affiliation{Physics Department, Nam\i k Kemal University, Faculty of Science and Letters, Tekirda\u g, Turkey}

\date{\today}
\begin{abstract}
We consider a $D$ dimensional Kasner type diagonal spacetime where metric functions depend only on a single coordinate and electromagnetic field shares the symmetries of spacetime.  These solutions can describe  static cylindrical or cosmological Einstein-Maxwell vacuum spacetimes. We mainly focus on electrovacuum solutions and four different types of solutions are obtained in which one of them has no four dimensional counterpart. We also consider the properties of the general solution corresponding to the exterior field of a charged line mass and discuss its several properties. Although it resembles the same form with four dimensional one, there is a  difference on the range of the solutions for fixed signs of the parameters.  General magnetic field vacuum solution are also briefly discussed, which reduces to Bonnor-Melvin magnetic universe for a special choice of the parameters. The Kasner forms of the general solution are also presented for the cylindrical or cosmological cases.
\end{abstract}
\maketitle

\section{Introduction}
\label{intro}
Higher dimensional extensions of general relativity  
are an active field of research. Although the existence of extra dimensions is still speculative, this topic plays an important role on the unification of the fundamental fields. For example, in Kaluza-Klein theory \cite{Kaluza,Klein}, electromagnetism and four dimensional gravity can be unified, with the expense of a scalar dilaton field, through a five dimensional pure gravity theory with a compact extra dimension having a small radius at the order of Planck length. In Braneworld models \cite{Brane,Antoniadis,Randall1,Randall2}, extra dimensions are considered as large, but all the matter fields are constrained to live in a three dimensional subspace called the three brane and only gravity can propagate into extra dimensions. Several exact solutions of extra dimensional theories were considered, where, in general, the focus is on the solutions having a horizon and which are asymptotically flat, i. e.  black hole type solutions, due to their striking properties.  (For a review see \cite{Emparan}). However, solutions which are not 
asymptotically 
flat or not having an event horizon, such as 
cylindrically symmetric ones, are studied less.

In four dimensions, cylindrically symmetric static vacuum and Einstein-Maxwell solutions were studied extensively.
The vacuum solution was found by Levi-Civita \cite{LC} whereas vacuum Einstein-Maxwell solutions were presented by several authors, such as Bonnor \cite{Bonnor} and Raychaudhuri \cite{Raychaudhuri} (See also \cite{LWitten,Safko,MacCallum,Richterek,Miguelote,BaykalEM,Stephani}). These solutions were often used for investigating some physically interesting problems, for example they can describe the exterior regions of static vacuum or current carrying cosmic strings \cite{Vilenkin,Superconducting,Moss,Puy}, which were believed to play important roles in structure formation of the universe. They are the only type of topological defects which are still compatible with recent observations, although their contribution must be less than $10\% $  in spectral power of CMB density fluctuations \cite{Pogosian,Bevis}. Recently, an interest emerged in higher dimensional generalizations of these strings \cite{Clement,Bronnikov} and other cylindrical sources \cite{Ponce,Baykal}. Moreover, one particular solution of cylindrical 
Einstein-Maxwell solutions, the Bonnor-Melvin magnetic universe \cite{BonnorM,Melvin,Thorne}, describing a uniform magnetic field along the symmetry axis is studied extensively in several contexts especially since it is possible to embed a  black hole in this magnetic field \cite{Ernst} via Harrison type transformations \cite{Harrison}. Its generalizations were studied in higher dimensional models \cite{Gibbons1,Gibbons2,Ortaggio1,Ortaggio2}. Some five dimensional Kaluza-Klein type solutions considering Melvin-type configurations were also studied \cite{Melvin5d,Das}. Thus, it might be important to obtain general static, cylindrical vacuum Einstein-Maxwell solutions to higher dimensions.

There are several motivations leading to this work. First of all, cylindrically symmetric solutions are one of the most important class of solutions of general relativity because they have a lot of applications on the topics such as gravitational waves \cite{EinsteinRosen}, gravitational collapse of non compact bodies and hoop conjecture, \cite{Collapse,Echeveria,ThorneKS,Thorne2}, cosmic strings \cite{Vilenkin,Superconducting,Moss,Puy}, quantum gravity \cite{Quantum,Kucher,Astekhar,Korotin}, numerical relativity \cite{Numerical}, etc. It is obvious that at least some of these topics are important for theories involving higher dimensions. Hence cylindrically symmetric solutions must also be discussed in higher dimensional theories, in order to complete the discussion of mathematical and physical properties of these theories. This motivates us to study all Einstein-Maxwell vacuum solutions of a static, cylindrically symmetric space time in higher dimensions. These solutions could be useful to understand some properties of the extended line-like configurations in higher dimensions similar to how four dimensional solutions  had been useful to understand four dimensional ones \cite{Vilenkin,Superconducting,Moss,Puy}. Indeed, it is known that these kind of solutions naturally produced in some theories employing higher dimensions. For example, in brane inflation theory, cosmic strings are produced during brane collisions  \cite{Tye,Tye2} with possible observable effects. Also, it was shown that cosmic $F$ and $D$ (super)strings \cite{Wittensuper} can be macroscopic  and also can become stable under some conditions \cite{dfstrings,Copeland}. This work could be seen as one of the first steps in the direction of understanding higher dimensional charged line like configurations. Hence, in this work, we only consider static fields and a diagonal metric ansatze similar to four dimensional static ones. More general solutions such as stationary ones or Einstein-Rosen type generalizations of these solutions or solutions in more generalized theories can be considered in further works.

 Our paper is organized as follows. In the next section we will first discuss the compatibility of the electromagnetic field sharing the symmetries of spacetime with the metric and show that the potential one form of the Maxwell field should have only one nontrivial component. Then by choosing this component as timelike, we obtain several possible general electrovacuum solutions for a cylindrically symmetric metric. We will also investigate some physical properties such as mass and charge per unit lenght for weak field limit and for general case. Moreover, some possible motions of charged or neutral test  particles for general solution corresponding to exterior of a charged infinite line mass are also discussed.  We also present vacuum solutions with a magnetic field. In the last section, using the property that cylindrical vacuum solutions can also be expressed in a Kasner type form, we present these general solutions and their cosmological counterpart in Kasner type coordinates.  

\section{Static Cylindrically Symetric Einstein-Maxwell  Solutions}
\subsection{Static Maxwell field and  Diagonal Spacetime}
In this paper we consider a $D$ dimensional diagonal spacetime where the metric functions depend only on one coordinate, let say $x^1$.   This spacetime can be written in the following Kasner-type diagonal form:
\begin{equation}\label{Kasmet}
ds^2=g_{\mu\nu}dx^\mu dx^\nu=-Y_0^2(x^1)dt^2+\sum_{i=1}^{D-1}\left[Y_i(x^1)dx^i\right]^2.
\end{equation}
 We are interested in  Einstein-Maxwell vacuum solutions of this spacetime. We consider a Maxwell field,
\begin{equation}
F=\frac{1}{2}F_{\mu\nu} dx^\mu\wedge dx^\nu
\end{equation}
sharing the symmetry of the metric, i. e., the vector potential is proportional to the Killing vectors of spacetime:
\begin{equation}
 \xi_\mu=\partial_\mu, \quad (\mu\neq 1).
\end{equation}
We consider a general vector potential one form:
\begin{equation}\label{Agen}
A=\sum_{\mu=0 (\mu \neq 1)}^{D-1}f_\mu (x^1)\, dx^\mu. 
\end{equation}
This yields that the Faraday tensor has only nonzero components as follows:
\begin{equation}
F_{1 \mu}=-F_{\mu 1}=f'_\mu(x^1),\quad (\mu\neq 1).
\end{equation} 
For this metric and Maxwell field, we want to solve the Einstein-Maxwell Equations 
\begin{eqnarray}
dF=d*F=0,\label{Maxwelleqn}\\
G_{\mu \nu}=\kappa\, T_{\mu \nu}, \label{Fieldeq} 
\end{eqnarray}
where the Maxwell Energy-momentum tensor is defined as
\begin{equation}
 T_{\mu \nu}=\frac{2}{\kappa} \left( F^{\lambda}_{\phantom{a}\mu}F_{\lambda \nu}-\frac{1}{4} g_{\mu \nu} F^{\lambda \kappa}F_{\lambda \kappa} \right).\label{Tmn}
\end{equation}

The Maxwell equations (\ref{Maxwelleqn}) yield
\begin{equation}
f_\mu(x^1)=\int \frac{C_{\mu} (Y_1 Y_\mu)^2}{Y}\, dx^1, 
\end{equation}
where $C_{\mu}$ are constants with $C_{1}=0$ and the function $Y$ is defined as
\[
Y \equiv \sqrt{-\mbox{det} g\,}=Y_0 Y_1\ldots Y_{D-1}.
\]
Trace of $T_{\mu\nu}$ yields
\begin{equation}
T=T^{\mu}_{\ \mu}=\frac{2(4-D)}{\kappa}F^{\mu\nu}F_{\mu \nu}=\frac{4-D}{\kappa}\sum_\mu {f'}_{\mu}^2 .
\end{equation}
The Einstein tensor of this metric is diagonal. However, the energy-momentum tensor  $T_{\mu\nu}$ has nondiagonal terms proportional to
\begin{equation}
T_{\mu\nu}\sim f'_\mu f'_\nu,\quad \mu\neq \nu.
\end{equation}
Hence, as in the four dimensions \cite{Miguelote}, the compability of the field equations demand the following cases:
\begin{itemize}
 \item[i)] Vanishing of Faraday tensor $F_{\mu\nu}$, i. e. vacuum solution,
 \item[ii)]Only one of $f_{\mu}$ is nonzero, i e., the vector potential is proportional to only one of the killing vectors $\xi_\mu$, either spacelike or timelike.
\end{itemize}
Thus, we have shown that the electromagnetic field one form should be proportional to one of the Killing vectors of spacetime, i. e. only one of $f_\mu$ in $(\ref{Agen})$ should have nonvanishing derivative. 
 
Depending on the choice that the vector potential one form has either timelike or spacelike nontrivial component, it describes an electric or magnetic field vacuum solution. Now we will solve the full Einstein-Maxwell field equations for a nonvanishing electrical field. The magnetic field solutions can be obtained by conveniently relabeling the coordinates.  

\subsection{Static Electrovacuum Solutions}

In order to solve the field equations, for convenience,  we consider a different metric ansatz. First we transform the metric (\ref{Kasmet}) such that after the transformation the $g_{11}$ and $g_{22}$ components have the same form, by choosing a new coordinate $\tilde{x}^1$ such that $Y_1 dx^1=\tilde{Y}_2 d\tilde{x}^1$.  Then we relabel the coordinates by  choosing $\tilde{x}^1=r, x^2=z, x^3=\phi$ and the metric functions $\tilde{Y}_0=e^U,\tilde{Y}_1=\tilde Y_2=e^{K-U}, \tilde Y_3=e^{-U}W,\tilde Y_i=X_i$.  Then we obtain a static, cylindrically symmetric, $D$ dimensional spacetime in Weyl type coordinates as follows \cite{Gibbons1,Gibbons2,Ponce}:
 \begin{eqnarray}
 ds^2&=&-e^{2U}dt^2+e^{2(K-U)}(dr^2+dz^2)+e^{-2U}W^2d\phi^2+\sum_{i=4}^{D-1}X_i^2(dx^i)^2,  \label{cylemet} 
\end{eqnarray}
where $K,U,W,$ and $X_i$ are the functions depending only on the radial coordinate $r$. Here the coordinates $t,r,z,\phi$ are usual cylindrical coordinates and the extra coordinates are labeled by $x^i, i=4,5,...,D-1$. Note that although the metric and field equations can be written more symmetrical by using Kasner type coordinates, we prefer this form since it will be easier to compare the solutions with four dimensional ones. Moreover,  Einstein-Rosen type coordinates can be obtained by transformation $t=iz,z=it$. This metric posesses in general $D-1$ Killing vectors $\xi_\mu= \partial_{\mu}, (\mu\neq r)$.
 Due to the discussion above, the potential one-form for the electromagnetic field of the spacetime (\ref{cylemet}) is chosen as
\begin{equation}\label{vectorpot}
A=f(r)\,dt,
\end{equation}
and the corresponding electromagnetic two form $F=dA$ becomes $F=f'dr\wedge dt$. The solutions for this configuration might be interpreted as $D$ dimensional vacuum Einstein-Maxwell solutions corresponding to 
exterior field of an infinitely long static charged line-mass with cylindrical symmetry.

The Einstein and Energy-Momentum tensors corresponding to the space-time and field configuration given by (\ref{cylemet}),(\ref{vectorpot}) are presented in ``Appendix'', where we have also shown that the field equations can be reduced to a very compact form. We will however employ a different strategy to obtain the solutions as follows.

Let us first define the functions $\Omega$ and $X$, where $X$ is actually the square root of the determinant of the extra dimensional part of the metric (\ref{cylemet}):
\begin{equation}
 \Omega=WX, \quad X=\prod_{i=4}^{D-1}X_i  . \label{OmegaX}
\end{equation}
Maxwell equations (\ref{Maxwelleqn}) for the metric (\ref{cylemet}) and the electromagnetic field (\ref{vectorpot}), with the help of above definition and (\ref{Amax} ) reduces to
\begin{equation}
f'=f_0 \frac{e^{2U}}{\Omega},
\end{equation}
where $f_0$ is a constant. Our strategy to find the solutions is to determine the unknown metric functions by eliminating $f'$ terms in the field equations (\ref{G00})-(\ref{Gii}) via adding or subtracting the equations one by one. If all the metric functions can be determined with this procedure, we can fed them into the Maxwell and Einstein equations to fully determine the Maxwell field $f$ as well.

Our first observation for solving these equations is that, when we add (\ref{G11})  and (\ref{G22})  we find
\begin{equation}
\Omega''=0 ,
\end{equation}
which imply that 
\begin{equation}
\Omega=W_0 r+W_1 \label{WX},
\end{equation}
with $W_0,W_1$ being constants.  We have to consider two different cases either $\Omega\sim r$ or $\Omega\sim \text{constant}$ by taking $W_1=0$ or $W_0=0$ respectively, since they may lead to different classes of solutions. Note that both of the  choices enable us to eliminate  one of the metric functions  $W,X_i$ in terms of others.

\subsubsection{$W_0\neq 0$ Case:}

For this case we can choose the constant $W_1=0$, thus we have $W=W_0 r/X$. We realize that subtracting any two of the extra dimensional field  equations  $(\ref{Gii}), (\ref{Tii})$ yield a second order differential equation involving only the functions corresponding to that coordinates. Namely,  if indices $i,j $ labels two of the extra dimensions, then this prodecure yields
\begin{equation}\label{XiXj}
 \left( r \frac{X_i'}{X_i}\right)'=\left(r \frac{X_j'}{X_j} \right)',
\end{equation}
which enables one to find all $X_i$'s in terms of one of the function $X_j$, $j$ fixed.
 Moreover, after using this fact, the field equations also permit us to find a differential equation involving the function $X_j$  and $U$ as follows:
\begin{equation}
 \left[r\left(\ln X_j\right)'\right]'=-\frac{1}{D-3}(rU')',\label{XkU}
\end{equation}
By integrating this equation, $X_j$ can be expressed in terms of $U$. Then we can easily find an equation, whose solution relates the function $K$ in terms of $U$,
\begin{eqnarray}
(r K')'=\frac{D-4}{D-3}(r U')'.\label{KU}
\end{eqnarray}
Remaining field equations can be integrated to find $U$. Note that the arbitrary integration constants can be fixed when these solutions fed into field equations again. By conveniently selecting the integration constants to have familiar forms, the metric (\ref{cylemet}) and field functions (\ref{vectorpot})  are found as,   

\begin{eqnarray}
K(r)&=&\left[2\sigma (2\sigma+p)+\frac{p^2+q^2}{2}-p\right] \ln r-\frac{D-4}{D-3}\ln(c_1 +c_2 r^{4\sigma}),\nonumber\\
U(r)&=&2\sigma \ln r-\ln(c_1+c_2 r^{4\sigma}),\nonumber\\
W(r)&=& W_0 r^{1-p}(c_1+c_2 r^{4\sigma})^{\frac{4-D}{D-3}}, \label{cylemetsol}\\
X_i(r)&=& r^{p_i} (c_1+c_2 r^{4\sigma})^{\frac{1}{D-3}},\nonumber\\
A&=&\sqrt{\frac{D-2}{D-3}}\sqrt{\frac{-c_1}{c_2}}\frac{1}{(c_1 +c_2r^{4\sigma})}dt, \label{cyleA} \\
p&=&\sum_{i=4}^{D-1} p_i,\quad q^2=\sum_{i=4}^{D-1} p_i^2.\label{pi}
\end{eqnarray}

Here, the positivity of the energy density requires the constants $c_1,c_2$  to satisfy
\begin{equation}\label{c1c2sigma}
 c_1 c_2\, \sigma^2<0.
\end{equation}
These solutions have free parameters $\sigma,W_0,p_i,c_1,c_2$ and since one of the constants $c_1$ or $c_2$ can be set to unity by rescaling the coordinates, we have $D-1$ free parameters. Further properties of this solution will be discussed in the next subsection.

Apart from the general solution discussed above, there is a special solution. For the solution of the equations (\ref{XiXj}),(\ref{XkU}),(\ref{KU}), if one sets all the first integration constants to zero, then one finds that $X_i(r)=e^{-\frac{1}{D-3}U(r)}$, $K(r)=\frac{D-4}{D-3} U(r)$ and $U(r)=-\ln[a\ln(c r)]$ and the metric and potential one form takes the form
\begin{eqnarray}
 ds^2&=&-\frac{dt^2}{(a \ln cr)^2}+ (a
\ln cr)^{\frac{2}{D-3}}\bigg(dr^2+dz^2+r^2d\phi^2 +\sum_idx_i^2 \bigg),\quad \\
A&=&\sqrt{\frac{D-2}{D-3}}\frac{1}{a \ln cr}dt,
\end{eqnarray}
where $a$ is a positive integration constant. This solution generalizes Raychaudhuri's solution \cite{Raychaudhuri} to higher dimensions. This conformastatic solution belongs to higher dimensional generalization of Papapetrou-Majumdar (PM) class\cite{Majumdar}. Note that there is a recent interest on higher dimensional generalizations of PM solutions \cite{MPhigh} .

\subsubsection{$W_0=0$ Case}

For the case $W_0=0$ in (\ref{WX}),
repeating similar steps we find that for this case there are also two different solutions. The first one has no four dimensional counterpart:
\begin{eqnarray}
&& X_i=e^{-\frac{U}{D-3}+p_i r},\label{Xisp}\\
&&K=\frac{D-2}{D-3}U+k_0 r,\\
&&W=\frac{W_1}{\prod_i X_i} ,\\
&&e^U=\frac{1}{\cos (f_0 r)},\\
&&f=\sqrt{\frac{D-2}{D-3}}\tan(f_0 r),\\
&& f_0=\sqrt{\frac{D-3}{D-2}}\sqrt{p^2+q^2},
\end{eqnarray}
where here $f_0$ is a constant, $p,q$ are defined in (\ref{pi}) and we have set some integration constants to zero.
 
The second solution of this class  is \begin{eqnarray}
ds^2&=&-\left(\frac{f_1}{r}\right)^2 dt^2+ \left(\frac{r}{f_1}\right)^{\frac{2}{D-3}}\bigg[e^{2 k_0 r}(dr^2+dz^2)+w_1^2d\phi^2+\sum_idx_i^2 \bigg],\quad\\
A&=&\sqrt{\frac{D-2}{D-3}}\frac{f_1}{r}\,dt,
\end{eqnarray}
with $f_1,k_0$ and $w_1$ are constants. This solution can be easily obtained if one sets all $p_i$ in (\ref{Xisp}) zero and then proceeds to the solution. 
The Kretchmann scalar of this solution is singular only at $r=0$, i. e.,
$R_{abcd}R^{abcd} \sim e^{-4(k_0r)}r^{-6}$. 
When $k_0=0$ this solution reduces to a special case of the general solution (\ref{cylemetsol})-(\ref{pi}) for $\sigma=1/2$ and $p_i=0$.

\subsection{Properties of the General Solution}

\subsubsection{The solution and its vacuum and Minkowski limits}
Hereafter we only consider the general solution given in (\ref{cylemetsol}), (\ref{cyleA}) and (\ref{pi}).
Let us recall the metric and electromagnetic field:
 \begin{eqnarray}
  ds^2&=&-\frac{dt^2}{G(r)^{2}}+G(r)^{\frac{2}{D-3}}
 r^\frac{4\sigma}{D-3}
 \bigg[ r^{4\sigma(2\sigma+p-1)-2p+p^2+q^2}\left(dr^2+dz^2 \right)\nonumber \\
&&\qquad \qquad \qquad \qquad \qquad +  \,W_0^2 r^{2(1-2\sigma-p)}d\phi^2 
+\sum_{i=4}^{D-1}r^{2p_i}(dx^i)^2 \bigg],\quad
  \label{cylemet1} \\
A&=& \frac{r}{P G} \frac{dG}{dr}\, dt, \label{cylemet1a}\\
G(r)&=&\left(c_1 r^{-2\sigma}+c_2 r^{2\sigma} \right) \label{cylemet1b},\\
P^2&=&-\frac{D-2}{D-3} 16\, c_1 c_2 \sigma^2. \label{P}
 \end{eqnarray}
Here we have written (\ref{cylemet1a}) in a slighly different form,  in order to compare the conventional form of four dimensional  solutions \cite{Stephani},  which differs from (\ref{cyleA}) by a constant term. We see that for $D=4$ the solution given above is in accordance with given in \cite{Stephani} up to notational  differences.

The Electromagnetic field two form has the expression
\begin{equation}
 F= \frac{P}{r\, G(r)^2}\, dt\wedge dr,\label{FEm}
\end{equation}
which vanishes when one of the parameters $c_1,c_2$ or $\sigma$ vanishes.
When we set $\sigma=0$ we obtain 
\begin{eqnarray}
ds^2&=&-\frac{dt^2}{(c_1+c_2)^2}+(c_1+c_2)^{\frac{2}{D-3}}\bigg[ r^{p^2+q^2-2p}(dr^2+dz^2)\nonumber \\
&& \qquad \qquad +\, W_0^2 r^{1-2p}d\phi^2+\sum_{i=4}^D r^{2p_i}(dx^k)^2  \bigg]. 
\end{eqnarray}
This is a vacuum solution but, unlike $D=4$, it is not locally flat in general. Hence setting $\sigma=0$ is not enough to obtain a locally flat metric, we also need  all the constants $p_k$ to vanish. The flat Minkowski limit of this metric requires the other constants of the solutions to satisfy $c_1+c_2=1$ and the conicity parameter $W_0=1$.

In order to find the relation of the corresponding Minkowskian solution of the Electromagnetic field of infinitely long charged line, we  expand (\ref{FEm}), series in $\sigma$.  We obtain, by also setting $c_1+c_2=1$ due to the discussion above,
\begin{equation}
F_{10}=E_r=\frac{2\lambda_0}{r}-8(c_2-c_1)\frac{\lambda_0 \sigma}{r}\log(r)+O(\sigma^3),
\end{equation}
where $\lambda_0$ can be seen as the charge per unit lenght of the Minkowskian solution
\begin{equation}\label{chargeperlambda0}
\lambda_0=-2\sigma\, \sqrt{\frac{D-2}{D-3}}\, \sqrt{-c_1c_2}. 
\end{equation}

Let us discuss the vacuum limits of this solution. As we have discussed above, when we set $\sigma=0$ we obtain a vacum solution and if remaining constants $p_i$ are also vanish we obtain (locally) flat solution.
For $c_1,\sigma\neq 0, c_2= 0$  we obtain higher dimensional generalization of static cylindrical vacuum solution, i.e., Levi-Civita solution.
Note that in this limit the constant  $c_1$ must be chosen as  positive in this case. By rescaling the ignorable coordinates and setting $c_1=1$ we can express this metric as follows:
\begin{eqnarray}\label{LChigher}
ds^2&=&-r^{4\sigma}dt^2+ \bigg[ r^{2(2\sigma+p)(2\sigma-1)+p^2+q^2}\left(dr^2+dz^2 \right)\nonumber \\
&&\qquad \qquad +\,W_0^2 r^{2(1-2\sigma-p)}d\phi^2 
+\sum_{k=4}^{D-1}r^{2p_k}(dx^k)^2 \bigg],
\end{eqnarray}
When we set $c_1=0,c_2,\sigma\neq 0$ to obtain vacuum solution from electrovacuum one, the electromagnetic field tensor vanishes identically, but  the vector potential becomes infinity. However, we can cure this ambiguity using the freedom to add a constant term to the vector potential which cancels the term causing this infinity, namely we can add the term  $-\{(D-2)/[(D-3)(-c_1 c_2)]\}^{1/2}$  to (\ref{cyleA}). 
Note that for this case ( $c_1=0, c_2\neq 0$)  in order to  obtain vacuum solution in the form  (\ref{LChigher}), we need to redefine  the parameters as $\sigma\rightarrow -\sigma$ and $p_i \rightarrow p_i-4\sigma/(D-3)$ with $c_2>0$.
 
 \subsubsection{The Singularity Structure and The Range of Solutions}
This spacetime, in general, contains two singularities, at  radii
\begin{equation}
r=0\quad  \text{and}\quad r=r_0=(-c_1/c_2)^{1/{4\sigma}}>0.\label{r_0}
\end{equation}
 This can be easily seen from scalars constructed  from curvature or field components, in which some of them are given below:
\begin{eqnarray}\label{F2}
F_{\mu \nu}F^{\mu \nu}&=&32\frac{D-2}{D-3}\frac{c_1c_2
\sigma^2\, r^{2(4\sigma+p) }}{r^{2(1+4\sigma^2+(p^2+q^2)/2+2\sigma p)}(c_1+c_2r^{4
\sigma})^{2\frac{D-2}{D-3}}}=-(E_1)^2
<0,\nonumber \\
F_{\mu\nu}*F^{\mu\nu}&=&0,\label{Fs2}\\
 R&=&16\bigg(\frac{D-4}{D-3}\bigg)\frac{c_1 c_2 \sigma^2\,  r^{2(4\sigma+p) } }{r^{2(1+4\sigma^2+(p^2+q^2)/2+2\sigma p)} (c_1+c_2 r^{4
\sigma} )^{2\frac{D-2}{D-3}}},  \nonumber \label{R2}
\end{eqnarray}
and similar expressions for $R_{\mu\nu}R^{\mu\nu}$ and $R_{\mu\nu\lambda\kappa}R^{\mu\nu\lambda\kappa}$.
The singularity at $r=r_0$ is a naked singularity since it is not surrounded by an event horizon. This singularity stems from the term   ($c_1+c_2 r^{4\sigma}$) in the denominator of the curvature scalars and becomes finite only for $\sigma=0$, i. e. only for vanishing of the electrical field.   The singularity at $r=0$ is also  a genuine singularity except for when the parameters satisfy the following equality $4\sigma+p=1+4\sigma^2+(p^2+q^2)/2+\sigma p $, including the special case $\sigma=1/2,p_i=0$,  in which the axis $r=0$ is not singular for this case. In this special case the coordinate $r$ must be extended to include $-\infty <r<r_0$. The interesting point of this value of parameters $\sigma=1/2,p_i=0$ is that it is the higher dimensional generalization of four dimensional case where the Levi-Civita metric describes an infinite  plane  geometry \cite{Gautreau,Miguelote,Arik} rather than a cylindrical one. Thus, for this case we have a charged infinite plane at $r=r_0$ where $r$ in this case should be 
considered as a Cartesian coordinate.

At $r\rightarrow \infty$ the solutions becomes asymptotically flat, since the leading terms in curvature scalars behave like $1/r^{h}, (h>0)$ which vanish as $r\rightarrow \infty$.   

An interesting point in these solutions is the choice of the sign of $c_1$ and $c_2$, since only the sign of their multiplication is fixed, i. e., $c_1c_2<0$.  For $D=4$, the function $G(r)$ can be positive or negative, since its square enters in the metric. For $D=5$, for the case $G(r)<0$ the signature of the metric becomes $-5$, i. e., it becomes an Euclidian metric with an overal minus sign, so this case is not physical. Moreover, for $D>5$, the regions in which $G(r)<0$ is also forbidden, since for these regions,  for $D>5$ the metric becomes  imaginary. Thus, although the solution is similar in its functional form for any $D\ge 4$ dimensions,  there is a big difference for four, five and higher dimensions due to the exponent $2/(D-3)$  on the metric function $G(r)$. The ranges in which metric is valid for $D>4$ can be classified as follows. Solution is valid for $r \in (r_0,\infty)$ whenewer $\sigma>0$, $c_1<0$, 
$c_2>0$ or $\sigma<0$, $c_1>0$, $c_2<0$, whereas the solution is valid for $r \in (0,r_0)$ for vice versa. Hence, unlike for $D=4$, choice of signs of $\sigma$, $c_1,c_2$ determines whether  the solution is valid only in `the interior' ( $0< r< r_0$) region or 'the exterior'  ($r_0<r<\infty$) regions. Only for $D=4$, the solution can be valid in both regions irrespective of which one of $c_1$, $c_2$ is negative.
 
Unlike $D=4$, since for $D\ge 5$  for the fixed signs of $c_1,c_2$ and $\sigma$, both the ``interior'' and exterior`` regions cannot be covered by the same metric and the physics in the interior region  is not clear since it is bounded by two singularities,  we consider the exterior region as the physically valid region of solution. Note that $\sigma>0$ and $\sigma<0$ cases are identical when we replace $c_1$ and $c_2$, thus we can only consider the case $\sigma>0$.  We can  actually set the location of $r_0$ as the axis of solution with a new radial coordinate $r'$. Then we have a solution well behaving at $r' \in (0,\infty)$ with a singularity at $r'=0$.

\subsubsection{Charge and Mass Per Unit Coordinate Length}

Let us calculate charge per unit coordinate length for the space time by using Gauss's teorem for the electrical flux across $r=\mbox{constant}$ surface \cite{Synge} 
\begin{equation}
\int_0^{2\pi}\int_0^1\ldots\int_0^1 F^{01}\sqrt{-g}\,dx^{D-1}\ldots dz\, d\phi=4\pi \lambda \label{gaussq}.
\end{equation}
where $g$
is the D-dimensional determinant of the metric and $\lambda$ is the parameter corresponding to the charge per unit coordinate length. Performing the integral (\ref{gaussq}) we obtain 
\begin{equation}
\lambda=-2\sigma\,  W_0\sqrt{\frac{D-2}{D-3}}\sqrt{-c_1c_2}.
\end{equation}
Note that this resut is in accordance with the weak field epression we have derived above, namely, charge per unit lenght of an infinitely long charged line in classical electrodynamics (\ref{chargeperlambda0}) since, in that limit we have to take $W_0=1$.

The mass per unit coordinate length at coordinate $r$ can be obtained by using the relativistic form of Gauss's teorem on gravitational flux  \cite{Whittaker} as follows
\begin{equation}
-\int_0^{2\pi}\int_0^1\ldots \int_0^1\frac{d^2r}{ds^2}\sqrt{-g}\,dx^{D-1}\ldots dz\,d\phi=4\pi\mu(r),\label{gaussm}
\end{equation}
where here $\mu(r)$ is the gravitational mass per unit coordinate length which depends on the coordinate $r$. For a neutral test particle initially at rest, 
the gravitational force is given by
\begin{eqnarray}
\frac{d^2r}{d\tau^2}=&&-U' e^{-2(K+U)}.
\end{eqnarray}
 Using expressions (\ref{cylemet}, \ref{cylemetsol}) we find
\begin{equation}\label{mu}
\mu(r)=W_0\sigma\frac{c_1-c_2r^{4\sigma}}{c_1+c_2r^{4\sigma}}=-W_0 \sigma \frac{\left(\frac{r}{r_0}\right)^{4\sigma}+1}{ \left(\frac{r}{r_0}\right)^{4\sigma}-1}. 
\end{equation}
For vacuum Levi-Civita solution \cite{LC}, the parameter $\sigma$ is related with the mass per unit length of the source generating cylindrical vacuum and when $\sigma>0$ the source attracts the surrounding test particles and for negative $\sigma$ it repells them \cite{Gautreau,Wang}. It is known for four dimensions \cite{Bonnorgrg} that the presence of radial electrical field dramatically changes this behaviour. Here we will show that this is indeed the case for $D>4$ as well.  
Since we are interested to the region $r>r_0$, irrespective of the sign of $\sigma$,  when electric field is present, i. e., $c_1,c_2\neq 0$,  $\mu$ is always negative. Also, 
 as $r$ goes to infinity  $\mu(r)\rightarrow - W_0 \sigma$ and for the uncharged case $\mu(r)\rightarrow W_0 \sigma$, therefore we can conclude that, as in four dimensional solution \cite{Bonnorgrg}, the effect of the radial electrical field is to change the sign of the gravitational mass. Thus similar conclusion can be made as four dimensional cases \cite{Miguelote,Bonnorgrg} that, unlike Newtonian physics, an infinily long charged line cannot be produced by physically acceptable sources. 

This fact can also be understood from the weak field expansion of the metric. If we series expand  the metric function $g_{00}$ in the mass parameter $\sigma$, then we obtain
\begin{equation}
g_{00}=-\left(c_1+c_2 \right)^{-2}\left[1+4 \sigma \frac{c_1-c_2}{c_1+c_2}\ln{r}+ 8\sigma^2 \frac{(c_1-c_2)^2-2 c_1c_2}{(c_1+c_2)^2}(\ln{r})^2 \right]+O(\sigma^3),
\end{equation}
where we can set without loss of generality $c_1+c_2=1$  for nearly Minkowskian spacetime. Here we consider the region $r_0<r<\infty$ where solution behaves well then we have $\sigma,c_2>0$, $c_1<0$ or $\sigma,c_2<0,c_1>0 .$  Remembering that in the Newtonian limit we have $g_{00}=-(1+2U)$, where $U$ is the Newtonian potential of the mass distribution, we see that for this case we have
\begin{equation}
U=2\sigma \,(c_1-c_2)\,\ln{r}. 
\end{equation}
This yields a radial force expresion,
\begin{equation}
F_r=-\frac{dU}{dr}=-\frac{2\sigma(c_1-c_2)}{r}, 
\end{equation}
which is repulsive since the quantity $\sigma(c_1-c_2)$, which can be considered as mass per unit length in Newtonian limit,  is always negative. These expressions are compatible with (\ref{mu}) in the weak field limit. When  electromagnetic field vanishes, i,e, $c_1=0, c_2>0$ or $c_2=0, c_1>0$ this force becomes attractive or repulsive depending on $\sigma$ is positive or negative. Note that for the limit $c_1$ vanishes we have to remember setting $\sigma\rightarrow -\sigma$, due to the discussion at the end of the Sect. 2.3.1.

\subsubsection{Equations of Motion of Test Particles}

Let us consider a test particle with charge  $e$ and mass $m$ in this field. The equations of motion can be derived from the expression 
\begin{equation}
\frac{d^2x^{\mu}}{d\tau^2}+\Gamma^{\mu}_{\beta\lambda}\frac{dx^{\beta}}{d\tau}\frac{dx^{\lambda}}{d\tau}=\frac{e}{m}F^{\mu}_{\ \nu}\frac{dx^{\nu}}{d\tau} , \label{eom}
\end{equation}
where $\tau$ is the proper time of the particle. The only nonzero components of electromagnetic field are obtained as
\begin{equation}
F^{10}=-F^{01}=\sqrt{\frac{D-2}{D-3}}\sqrt{-c_1 c_2}\frac{4\sigma}{r}e^{2(U-K)}.
\end{equation}
The first integrals of the equations of motion can be calculated as follows:
\begin{eqnarray}
&&e^{2U}\frac{dt}{d\tau}=\frac{e}{m}f(r)-E \label{energy},\\
&&e^{2(K-U)}\,\frac{dz}{d\tau}=L,\label{Lmom}\\
&&e^{-2U}W^2\,\frac{d\phi}{d\tau}=J,\label{Amom}\\
&&X_i^2\frac{dx_i}{d\tau}=P_i,\label{Pmom}
\end{eqnarray}
where the integration constants $E,L,J,P_i$ are related with total energy, linear momentum in $z$ direction, angular momentum, and linear momentums in $x_i$ directions of the particle.

We have studied the equations of motion (\ref{eom}) in detail. Here  we only state some important results for a test particle (outside the singularity at $r=r_0$ for electro-vacuum solutions, i. e. $c_1<0$ for $\sigma>0$ and $c_2<0$ for $\sigma<0$):
\begin{itemize}
 \item When the electrical field presents, the charged and neutral test particles cannot follow a trajectory which is purely circular, axial, or in one of the $x_i$  directions.
\item For vacuum solution, i. e. when electrical field vanishes, test particles can follow circular timelike geodesics for $0<\sigma+p/4<1/4$ and circular null geodesics for $\sigma+p/4=1/4$. Such particle can also follow axial timelike (null) geodesics if the parameters satisfy the inequality (equality) $2\sigma(2\sigma -2+p)-p+(p^2+q^2)/2>0(=0)$.  These are in accordance with four dimensional Levi-Civita solution \cite{LC,Gautreau,Wang,Arik}.
\item For vacuum solution, a test particle  initially at rest experiences an attractive force for $\sigma>0$ and repulsive force for $\sigma<0$ in accordance with four dimensional solution \cite{LC,Gautreau,Wang,Arik}.
\item For electro-vacuum solution, neutral test particles initially at rest experience a repulsive force, for any value of parameters, as in four dimensional solution \cite{Bonnorgrg}.
\item The radial force on a charged test particle initially at rest in electro-vacuum solution (outside the singularity at $r_0$) is affected by two forces, in accordance with four dimensional case \cite{Bonnorgrg}, given by 
\begin{eqnarray}
\frac{d^2r}{d\tau^2}&\approx&\left\{ 
\frac{\lambda e}{m}\frac{r^{4\sigma}}{W_0} -\sigma c_1 \left[ \left(\frac{r}{r_0} \right)^{2\sigma}+1 \right]\right\}, \label{radforce}
\end{eqnarray}
where a positive proportionality factor is omited for clarity. Here, the second term in the first pharanthesis is the gravitational force, and it is always repulsive since always $c_1 \sigma<0$. The first term  however, the electrical force term, is attractive if the charge of the particle, $e$, and charge per unit length of the source, $\lambda$, have different signs and repulsive it both have same sign. When the electrical force on the particle is attractive, there are values of $r$ in which the particle is in neutral equilibrium where the gravitational and electrical forces cancel each other and the expression (\ref{radforce}) vanishes.  
 
\end{itemize}

\subsection{Cylindrical Vacuum Solutions with a Magnetic Field}

Here we present the general solution of the Einstein-Maxwell vacuum magnetic field solutions presented in 
Einstein-Rosen type coordinates. The solutions can be obtained by directly solving the field equations as we have done. They can also be obtained by suitably relabeling the coordinates and parameters as well. 
Let us present the resulting solution, for an electromagnetic potential one form  $A=h(r)d\phi$, compatible with the form given in \cite{Stephani}:
\begin{eqnarray}\label{magneticsoln}
 ds^2=&&G^{-2} W_0^2\, d\phi^2+G^{\frac{2}{D-3}}
r^\frac{2m}{D-3}
\bigg[ r^{2m(m+p-1)-2p+p^2+q^2}\left(dr^2-dt^2 \right)\nonumber \\
&&\qquad \qquad \qquad +\, r^{2(1-m-p)}dz^2 +\sum_{k=4}^{D-1}r^{2p_k}(dx^k)^2 \bigg], \\
&&G= \left(c_3\, r^{-m}+c_4\, r^{m} \right),\quad p=\sum_{i=4}^{D-1}p_i, \quad q^2=\sum_{i=4}^{D-1}p_i^2, 
 \\
A&=&\sqrt{\frac{D-2}{D-3}}\sqrt{\frac{c_3}{c_4}}\frac{r^{-m}}{G(r)}d\phi. \label{magneticvec}
\end{eqnarray}

 The positivity of energy density  for this case requires the condition
\begin{equation}\label{c3c4}
c_3 c_4\, m^2 >0.
\end{equation}
Vacuum limit is obtained when one of the constants $c_3,c_4$ is vanishing. One of $c_3,c_4$ can be set to unity, thus the solution has $D-1$ free parameters: $W_0,m,$ one of $c_3,c_4$ and $p_k$.  
Note that a magnetic field solution whose vector potential is in any one of the other spacelike Killing directions can be obtained with appropriate coordinate re-labelings. The common property of these magnetic field solutions, unlike electrical field solutions, that there is no singularity other than at $r=0$, despite the scalars constructed from field or curvature components have the same form with (\ref{Fs2}), due to the relation (\ref{c3c4}). For this solution the motion of charged and neutral test particles are discussed in detail in \cite{LWitten} for four dimensions, and for higher dimensional case similar behaviour can be expected for such particles. 

 For the special case $m=1,$ $p_i=0$, the solution reduces to higher dimensional generalizations of one of the important Einstein-Maxwell solutions, namely, Bonnor-Melvin magnetic universe \cite{BonnorM,Melvin,Thorne},
\begin{eqnarray}\label{magneticsolnMelvin}
 ds^2&=&\frac{W_0^2 r^2 d\phi^2}{\left(c_3 +c_4 r^{2} \right)^2}+\left(c_3 +c_4 r^{2} \right)^{\frac{2}{D-3}}
\left[-dt^2+dr^2+dz^2 
 +\sum_{k=4}^{D-1}(dx^k)^2 \right],\
 \\
A&=&\sqrt{\frac{D-2}{D-3}}\sqrt{\frac{c_3}{c_4}}\frac{1}{(c_3 +c_4\, r^{2})}d\phi. \label{magneticvecmelvin}
\end{eqnarray}
This solution describes a uniform 
magnetic field, i. e., Melvin flux-brane. This solution can be considered as a union of parallel magnetic flux lines extending along $z$ axis.  This spacetime is stable and it has no singularities. One important aspect of this solution is that, the flat Minkowski metric in square pharanthesis can be replaced with any $D-1$ dimensional Ricci flat {\it seed} solution \cite{Ernst}, which is a consequence of Harrison transformations \cite{Harrison} which enables one to generate an Electrovacuum solution from a vacuum Einstein solution. Several applications of this observation had been  made such as, a solution describing a back hole in an external magnetic field \cite{Ernst,Gibbons1,Gibbons2,Ortaggio1,Ortaggio2}.

\section{$D$ dimensional Kasner-type Einstein-Maxwell Solutions}


One of the earlier and important solutions of Einstein equations is Kasner solution \cite{Kasner}. Although it  is generally considered as a cosmological solution and related with Bianchi Type I  cosmological vacuum, its original form  was presented with a positive signature, i. e.,
\begin{equation}\label{kas4d}
ds^2=\sum_{k=1}^4(x^4)^{2a_k}(dx^k)^2,
\end{equation}
where metric  depends only on $x^4$  and  the parameters $a_i$  satisfy the following conditions:
\begin{equation}
a_1+a_2+a_3=1+a_4,\quad a_1^2+a_2^2+a_3^2=(1+a_4)^2. \label{kas4}
\end{equation}
Here the parameter $a_4$ can be set to zero without loss of generality.

Actually, homogeneous and anisotropic cosmological Kasner vacuum metric is a metric of the form given above with the timelike coordinate $x^4=t$ having $(-+++)$ signature
\begin{equation}
ds^2=-dt^2+\sum_{i=1}^3 t^{2a_i}(dx^i)^2,
\end{equation}
with the same conditions (\ref{kas4}) with $a_4=0$.  Note that if one choses the coordinate $x^4$ as the radial coordinate and the remaining coordinates as usual cylindrical coordinates in (\ref{kas4d}) with properly chosen signature, then one obtains the cylindrical static vacuum Levi-Civita solution \cite{LC} in its Kasner form. There is a simple transformation between these two different forms of cylindrical vacuum solutions (See for example \cite{Arik,Ponce}). Actually, Kasner type cosmological \cite{Chodos,Kokarev,Hervik} or cylindrical vacuum solutions \cite{Ponce,Baykal} were generalized to higher dimensions. Hence it is logical to  bring the obtained Einstein-Maxwell type solutions in this form.  Thus,  in this section, for completeness,  we present $D$ dimensional general electrical and magnetic field solutions we have discussed in the previous sections, namely cylindrical static Einstein-Maxwell solutions in generic dimensions, in Kasner type form. Moreover, this analogy enables us to present an anisotropic cosmological Kasner solution with a nonvanishing electromagnetic field. Let us first discuss  vacuum solutions in this form. 

\subsection{Vacuum Solutions}
The vacuum solution of the metric (\ref{Kasmet}) yields a multidimensional generalization \cite{Chodos,Kokarev,Hervik,Ponce} of Kasner solution  
\begin{equation}
Y_\mu=\tilde{C}_\mu\, x^{\alpha_\mu},\ \ \mbox{(no sum)},
\end{equation}
where $\tilde{C}_\mu$ and $\alpha_\mu$ are integration constants and for clarity we have taken  $x^1=x$. The $D$ dimensional Kasner parameters $\alpha_\mu$  satisfy
\begin{eqnarray}
 &&\sum_{\mu (\neq 1)=0}^{D-1} \alpha_\mu=1+\alpha_1, \label{kas1} \\
&&\sum_{\mu (\neq 1)=0}^{D-1} \alpha_\mu^2=(1+\alpha_1)^2. \label{kas2}
\end{eqnarray}
This metric describes a $D$ dimensional cylindrically symmetric static vacuum solution \cite{Ponce} if one identifies the coordinates $x^\mu$ as usual cylindrical coordinates, namely, $(x^0,x^1,x^2,x^3,x^i)=(t,r,z,\phi,x^i)$. Note that the constant $\alpha_1$ can be set to zero without loss of generality by defining a Gaussian normal coordinate $d\tilde{x}=x^{\alpha_1} dx^1$ and redefining other parameters. Hence, only  $D-3$ of the parameters $\alpha_\mu$ are free. These parameters define $D-2$ dimensional ``Kasner sphere'' analogous to the Kasner circle in four dimensions. This solution has a naked singularity at $x=0$ unless the parameters correspond  to locally flat, Rindler type, spacetime, e. g. only one of them is nonzero and equal to $1+\alpha_1$.

\subsection{ Electrovacuum Solutions}
For the electro-vacuum solution only  $A=f_0(x) dt$ term is present, describing an Einstein-Maxwell electrovacuum solution with a nonvanishing electrical field. As we have demonstrated in Section (2) that there are four different solutions. Here,
 we only present the Kasner form of the general solution (\ref{cylemetsol}), (\ref{cyleA}), (\ref{pi}), in the form of the metric (\ref{Kasmet}) and field (\ref{Agen}) as follows: 
\begin{eqnarray}\label{electrovacKas1}
  Y_0&=&(c_1 x^{\alpha_0}+c_2 x^{-\alpha_0})^{-1},\\
 Y_i&=&(c_1 x^{\alpha_0}+c_2 x^{-\alpha_0})^{1/{(D-3)}}x^{\alpha_i+\alpha_0/{(D-3)}}, \label{electrovacKas2}\\
f_0(x)&=&\sqrt{\frac{D-2}{D-3}}\,\sqrt{-\frac{c_1}{c_2}}\frac{1}{c_1+c_2 x^{2\alpha_0}},\\ 
 c_1 c_2 \alpha_0^2&<& 0, \label{electrovacKas3}\ \
 \end{eqnarray}
where again we have set $x^1=x$ for clarity and the parameters $\alpha_\mu$ satisfy (\ref{kas1},\ref{kas2}). The coordinates should be thought as cylindrical coordinates as in the previous subsection. Vacuum solution is recovered when $c_1=0$ or $c_2=0$.

The transformation from Weyl type coordinates (\ref{cylemet}) to Kasner type coordinates (\ref{Kasmet}) (with $\alpha_1=0$) are given by
\begin{eqnarray}
&& x=r^\Sigma,\ \alpha_0=\frac{2\sigma}{\Sigma},\ \alpha_2=1-\frac{1}{\Sigma}, \ \alpha_3=\frac{1-2\sigma-p}{\Sigma}, \nonumber \\
&& \alpha_{i(i=4,5,...,D-1)}=\frac{p_i}{\Sigma},\  p=\sum_{I=4}^{D-1} p_i, q=\sum_{I=4}^{D-1} p_i^2, \\ 
&& \Sigma=2\sigma (2\sigma+p-1)+\frac{p^2+q^2}{2}-p+1. \nonumber
\end{eqnarray}

\subsection{Magnetic Field Vacuum Solutions}

For only nonvanishing Maxwell field term is spacelike, i. e.  $F_{i1}=f'_i$ where fixed indice $i$ is representing the coordinate of any one of the space-like Killing vectors of spacetime, we have magnetic field solution with $A=f_i(x)dx^i$ (no sum).  
Since the special solutions can be expressed accordingly, let us only present the general solution (\ref{magneticsoln}), (\ref{magneticvec}) in Kasner form
\begin{eqnarray}
  Y_i&=&(c_3 x^{\alpha_i}+c_4 x^{-\alpha_i})^{-1},\\
 Y_{\mu (\mu\neq i)}&=&(c_3 x^{\alpha_i}+c_4 x^{-\alpha_i})^{1/{(D-3)}}x^{\alpha_\mu+\alpha_i/{(D-3)}}\\
f_i(x)&=&\sqrt{\frac{D-2}{D-3}}\,\sqrt{\frac{c_3}{c_4}}\frac{1}{c_3+c_4 x^{2\alpha_i}},\\  c_3 c_4 \alpha_i^2&>&0,
 \end{eqnarray}
where again the parameters to satisfy (\ref{kas1},\ref{kas2}) and the coordinates should be considered as cylindrical ones. Vacuum limit of this solution is when $c_3=0$ or $c_4=0$.

\subsection{Kasner-Maxwell Cosmological Solutions}

A generalization of Kasner cosmological vacuum solution with an electromagnetic field aligned in certain direction in four dimensions was given by Datta \cite{Datta}. The solution we have presented in the previous section can also be seen as a higher dimensional generalization of this solution, if one considers the coordinates as Cartesian coordinates and chooses the nonignorable coordinate as timelike. Then the solution  for a vector potential aligned along a fixed $x^i$ direction, i. e. $A=f(t)\,dx^i$ reads
\begin{eqnarray}
ds^2&=&\frac{(dx^i)^2}{\left(k_1 t^{\alpha_i}+k_2 t^{-\alpha_i}\right)^2}+\left(k_1 t^{\alpha_i}+k_2 t^{-\alpha_i}\right)^{\frac{2}{D-3}}
\sum_{\mu, \nu(\neq i)}\left[t^{2\alpha_\mu+\frac{2\alpha_i}{D-3}}\eta_{\mu\nu} dx^\mu dx^\nu \right],\nonumber\\
&&A=\sqrt{\frac{D-2}{D-3}}\,\sqrt{\frac{k_2}{k_1}}\frac{1}{k_2+k_1 t^{2\alpha_i}} dx^i,
\end{eqnarray}
where the Kasner  parameters $\alpha^\mu$ satisfy (\ref{kas1},\ref{kas2}) and the paremeters $k_1,k_2$  have to satisfy the condition
\begin{eqnarray}
k_1 k_2 \alpha_i^2>0. 
\end{eqnarray}
This spacetime is singular at $t=0$ in general. However, for $\alpha_i=1, \alpha_{\mu, (\mu\neq i)}=0$, similar to Melvin solution, it describes a time dependent electromagnetic field where spacetime is free of singularities.

\section{Discussion}

In this paper we have presented higher dimensional generalizations of Einstein-Maxwell vacuum solutions for diagonal metrics depending only on a single coordinate and corresponding to cylindrical or Kasner type spacetimes. We mainly considered the electro-vacuum solutions in cylindrical coordinates, since magnetic field solutions are easy to obtain if electrical one is known. We have found four different solutions in which one of them has no four dimensional analogue.  As an application, we have studied some properties of the general solution corresponding to the exterior field of a charged line mass. The properties of this solution does not differ from four dimensional one, i. e.  there are two singularities in general, one at the symmetry axis and the other at a certain distance from the axis. Also, as in four dimensions, the presence of the radial electrical field changes the characteristics of spacetime such that the mass per unit length becomes negative. However, there is an important distinction, in which, unlike four dimensional case, if the signs of the parameters satisfying the relation (\ref{c1c2sigma}) is fixed, then the solution is valid either at the region outside of the outer singularity or at the region in between the singularities but cannot be valid in both regions. We have also realized that as in four dimensional solution, the mass parameter becomes negative irrespective of the sign of parameters. This enables us to conclude that, as in the four dimensions \cite{Bonnorgrg}, a charged infinite line mass has not realistic counterpart in higher dimensional relativistic theory.  

These solutions may have interesting applications. For example as in the four dimensions, these solutions may describe the exterior regions of (space-like or time-like) current carrying (cosmic) strings, hence the parameters can be related to mass per unit length and  the four-charge density of these strings. For this interpretation, one has to consider a higher dimensional Nielsen-Olesen type Abelian Higgs model with appropriate gauge fields and discuss its asymptotic behavior of the spacetime considered in this model.

Another application is to consider a brane embedded in these spacetimes. Since the only geodesically complete spacetime in these solutions is the Melvin case, using the other cases seems to have problems since they contain naked singularities. For the electrical case, although there is a singularity at $r=0$ and $r=r_s$, if a brane located at $r\ge r_s$, since  a neutral brane will feel a repulsive force in this spacetime, it will repel from the singularity. Hence it seems the singularity may not as harmful as it thought. However these discussions need more detailed calculations, in which we leave for  future works. 

\section*{Appendix: Field Equations}

The nonzero components of the Einstein tensor for Einstein Maxwell field equations $G_{\mu\nu}=\kappa T_{\mu\nu}$ for the metric (\ref{cylemet})  are given by
\begin{eqnarray}
G^{0}_{\phantom{a}0}&=&-e^{2(U-K)}\left\{ 2U''-K''-U'^2+2U'\frac{W'}{W}-\frac{W''}{W} \right.\nonumber \\
&&\qquad \qquad \qquad  \left. +\, \sum_i^N\left[\left(U'-\frac{W'}{W}-\frac{1}{2}\sum_{j\neq i}^N \frac{X_j'}{X_j} \right)\frac{X_i'}{X_i} -\frac{X_i^{''}}{X_i} \right]   \right\},
 \label{G00}\ \ \\
G^{1}_{\phantom{1}1}&=& e^{2(U-K)} \left[ K'\frac{W'}{W}-U'^2+ \sum_i^N \left(K'-U'+\frac{W'}{W}+\frac{1}{2}\sum_{j\neq i}^N \frac{X_j'}{X_j} \right)\frac{X_i'}{X_i}  \right], \ 
\label{G11} 
\end{eqnarray}

\begin{eqnarray}
G^{2}_{\phantom{2}2}&=&e^{2(U-K)} \left\{U'^2-K'\frac{W'}{W}+\frac{W''}{W} \right. \nonumber \\ && \qquad \qquad \left.  + \sum_i^N\left[\left(U'-K'+\frac{W'}{W} +\frac{1}{2}\sum_{j\neq i}^N \frac{X_j'}{X_j}\right)\frac{X_i'}{X_i} +\frac{X_i^{''}}{X_i} \right]\right\}, \label{G22} 
\\
G^{3}_{\phantom{a}3}&=& e^{2(U-K)} \left\{ K''+U'^2 + \sum_i^N\left[\left(U'+\frac{1}{2}\sum_{j\neq i}^N \frac{X_j'}{X_j}\right)\frac{X_i'}{X_i} +\frac{X_i^{''}}{X_i} \right]  \right\}, \label{G33} \\
 G^{i}_{\phantom{i}i}&=&e^{2(U-K)}\left[ U'^2-U''+K''-\frac{U'W'}{W}+\frac{\left(W'\prod_{j\neq i}^N X_j \right)' }{W \prod_{j\neq i}^N X_j} \right. \nonumber\\
&& \qquad \qquad \qquad \qquad \qquad \qquad \qquad \left. +\sum_{j\neq i}^N\frac{X_j''}{X_j}+\frac{1}{2}\sum_{\substack{j,k\\i\neq j\neq k}}\frac{X_j'X_k'}{X_jX_k} \right], \label{Gii}  
\end{eqnarray}
and  the nonzero components of energy momentum tensor for and electric field (\ref{vectorpot}) is given by
\begin{eqnarray}
-T^{0}_{\phantom{0}0}=-T^{1}_{\phantom{1}1}=T^{2}_{\phantom{2}2}=T^{3}_{\phantom{3}3}=T^i_{\phantom{i}i}=\frac{1}{2}e^{-2K}f'^2. \label{Tii}
\end{eqnarray}

Let us use the functions $\Omega$ and $X$, defined in (\ref{OmegaX}).
From (\ref{G11}) and (\ref{G22}) we have
\begin{equation}\label{Omegadp}
\Omega''=0.
\end{equation}
Also from (\ref{G00}) and (\ref{G33}) 
\begin{equation}
\frac{(U'\Omega)'}{\Omega}=\frac{(W'X)'}{2X}+\frac{e^{-2K}\, f'^2}{2}.
\end{equation}
The Maxwell equations yield
\begin{equation}\label{Amax}
 \left( f'\Omega\right)'=2\Omega f'U'.
\end{equation}

From (\ref{G11})
\begin{eqnarray}
 K'\frac{\Omega'}{\Omega}&=&U'^2+U'\frac{X'}{X}-\frac{e^{-2K}\, f'^2}{2} \nonumber \\
&&+ \, \frac{1}{2}\left[\left(\frac{\Omega'}{\Omega}\right)^2-\left(\frac{W'}{W}\right)^2-\sum_i \left(\frac{X'_i}{X_i} \right)^2 \right],\ 
\end{eqnarray}
and from (\ref{Gii}) and (\ref{G00}) we have $D-4$ equations
\begin{equation}
 \left(\frac{\Omega X_i'}{X_i}\right)'= (\Omega U')'-\Omega f'^2 e^{-2K}.\label{Xiapp}
\end{equation}
The equations (\ref{OmegaX}), (\ref{Omegadp})-(\ref{Xiapp}) constitute a complete set and they are in accordance with the equations for $D=4$ case given in \cite{Stephani}.

\begin{acknowledgments}
We thank the anonymous referees for their usefull comments and suggestions. 
\end{acknowledgments}

\end{document}